\RequirePackage{lineno}
\documentclass[aps,prl,twocolumn,superscriptaddress,groupedaddress,preprintnumbers]{revtex4-2}  
\usepackage{graphicx}  
\usepackage{dcolumn}   
\usepackage{bm}        
\usepackage{amssymb}   
\usepackage{multirow}
\usepackage{epsfig}
\usepackage{amsmath}
\usepackage[colorlinks, linkcolor=blue]{hyperref}
\usepackage{ulem}

\newcommand{\effstw}{\ensuremath{\sin^2\theta_{\text{eff}}^{\text{$\ell$}}}}

\begin{document}

\lefthyphenmin=2
\righthyphenmin=2

\widetext

\title{ Factorization of the forward-backward asymmetry and measurements of the weak mixing angle and proton structure at hadron colliders }
\affiliation{Department of Modern Physics, University of Science and Technology of China, Jinzhai Road 96, Hefei, Anhui 230026, China}
\affiliation{Department of Physics, College of Sciences, Northeastern University, Shenyang 110819, China}
\affiliation{Department of Physics and Astronomy, Michigan State University, East Lansing, MI 48823, USA}

\author{Siqi Yang} \affiliation{Department of Modern Physics, University of Science and Technology of China, Jinzhai Road 96, Hefei, Anhui 230026, China}
\author{Yao Fu} \affiliation{Department of Modern Physics, University of Science and Technology of China, Jinzhai Road 96, Hefei, Anhui 230026, China}
\author{Minghui Liu} \affiliation{Department of Modern Physics, University of Science and Technology of China, Jinzhai Road 96, Hefei, Anhui 230026, China}
\author{Liang Han} \affiliation{Department of Modern Physics, University of Science and Technology of China, Jinzhai Road 96, Hefei, Anhui 230026, China}
\author{Tie-Jiun Hou} \affiliation{Department of Physics, College of Sciences, Northeastern University, Shenyang 110819, China}
\author{C.-P. Yuan} \affiliation{Department of Physics and Astronomy, Michigan State University, East Lansing, MI 48823, USA}

\begin{abstract}
The forward-backward charge asymmetry ($A_{FB}$) at hadron colliders is sensitive to both the electroweak (EW) symmetry breaking represented by 
the effective weak mixing angle $\effstw$, and the proton structure information in the initial state modeled by the parton distribution functions 
(PDFs). 
Due to their strong correlation, the precisions of determination on $\effstw$ and PDFs using the measured $A_{FB}$ spectrum are limited. 
In this paper, we define a set of structure parameters which factorize the unique proton information of the relative 
difference between quarks and antiquarks in the $A_{FB}$ observation. 
Other than the conventional way of extracting $\effstw$ from the convolution 
of PDF and EW calculations, we propose a new method to simultaneously determine the value of 
$\effstw$ and the proton structure terms by fitting to the observed $A_{FB}$ distribution, and point out 
the necessity of specifying additional observations to further reduce the uncertainties on the proton structure terms respectively, 
so that model-independent high precision $\effstw$ and proton structure measurements can be achieved 
at the future LHC experiments.

\end{abstract}
\maketitle

\section{Introduction}

The forward-backward asymmetry ($A_{FB}$) of the neutral current process $f_i \bar{f}_i \rightarrow Z/\gamma^* \rightarrow f_j \bar{f}_j$ 
describes the relative difference between cross sections of the forward and backward scattering. It arises from the vector and 
axial vector couplings of $Z$ to fermions, and is governed by the effective weak mixing angle $\effstw$, which has been measured 
first in the electron-positron collisions at the LEP and the SLC~\cite{LEP-SLD}, then in the proton-antiproton collisions at 
the Tevatron~\cite{TevatronStw}. At the Large Hadron Collider (LHC), the initial fermions and antifermions of the corresponding 
Drell-Yan (DY) process $pp(q\bar{q})\rightarrow Z/\gamma^* \rightarrow \ell^+\ell^-$ are quarks and antiquarks, acting as partons 
in the proton, with their momentum distributions modeled by the parton distribution functions (PDFs). 
Since quarks and antiquarks would come from either side of the proton beams, the observed asymmetry at hadron colliders, denoted 
as $A^h_{FB}$, is diluted from its original $A_{FB}$ value. The reduction in magnitude from $A_{FB}$ to $A^h_{FB}$ directly reflects 
the difference between quark and antiquark momentum fractions in the proton, and can be used to constrain PDFs. Some of the earlier 
studies of using $A^h_{FB}$ to constrain PDFs can be found in Refs.~\cite{ArieStudy, ATLASStudy1, ATLASStudy2, CPCstudy}.

Though $A^h_{FB}$ can be precisely measured at the LHC, 
it is difficult to use this observable to constrain the PDFs and extract a more precise value of $\effstw$, due to the strong correlation 
between them. 
On the one hand, as noted in Ref.~\cite{CPCstudy}, directly using the $A^h_{FB}$ observable in a typical global PDF analysis 
may yield a large bias induced by the limited precision of $\effstw$, which should be considered as an additional uncertainty 
in constraining the PDFs. 
On the other hand,  
as reported by the ATLAS, CMS and LHCb collaborations~\cite{ATLASstw, CMSstw, LHCbstw}, the PDF-induced uncertainty 
is becoming the most dominant issue in the $\effstw$ measurements, and thus
prevents the measurements of $\effstw$ at the LHC to achieve a relative precision at $\mathcal{O}(10^{-4})$.
Such precision is recently achieved in the $W$ boson mass measurement by the CDF collaboration~\cite{CDFWmass}, which is significantly 
deviated from the Standard Model (SM) prediction. If such deviation hints new physics beyond the SM, it should also appear 
as a shift on $\effstw$. 
Therefore, it is essential to have a strategy to decouple the parton information from the EW prediction in the $A^h_{FB}$ 
observation, so that $\effstw$ and parton information can be simultaneously determined using the LHC data instead of 
fitting for one and fixing the other in the conventional methods. 

However, such simultaneous determination is challenging 
in the framework of PDF global analysis, with $\effstw$ as a free parameter. Many of the old experimental data sets included in the PDF 
analyses were performed at the leading order in EW interaction, and those data are still important for providing PDF information~\cite{OldData}. 
To achieve a high precision of $\effstw$ determination, all the Wilson coefficients of the scattering processes included in the PDF analysis 
have to be updated to higher-order in EW interactions. This feature is not yet available in the current PDF global analysis programs. 

In this paper, we propose a new method which factorizes the proton information relevant to the $A^h_{FB}$ measurement into a set of 
structure parameters. They are defined to be generally independent of any specific PDF modeling, and thus can be viewed as some new 
experimental observables. These parameters, together with $\effstw$, can be determined from one single measurement of the $A^h_{FB}$ 
at the LHC, through a carefully-designed fitting procedure, as to be described below. The correlation between the determination of 
$\effstw$ and proton structure is automatically taken into account in the proposed analysis.

\section{Factorization of $A^h_{FB}$}

The $A_{FB}$ at hadron colliders is observed in the Collins-Soper (CS) frame~\cite{cs-frame}, which is a center-of-mass frame with 
the $\hat{z}$-axis defined as the bisector of the angle formed by the direction of the momentum of one incoming hadron ($H_A$) and 
the negative direction of the other incoming hadron ($H_B$). To decouple the original asymmetry from the proton structure information, 
we define two scattering angles, {\it i.e.} $\cos\theta_q$ and $\cos\theta_h$, with different choices of $H_A$ and $H_B$. 
For the $\cos\theta_q$, $H_A$ and $H_B$ are assigned according to the directions of $q$ and $\bar{q}$, respectively. 
The differential cross section of the DY process can be expressed in term of $\cos\theta_q$ as:

{\footnotesize 
\begin{eqnarray}
\label{eq01:dsigmadq}
\frac{d\sigma}{d\cos\theta_q dY dM dQ_T} &=&\frac{3}{8} \sum_f \alpha_f(Y, M, Q_T) \times \Big\{ (1+\cos^2\theta_q) \nonumber \\
&+& \frac{1}{2}A^f_0(Y, M, Q_T)(1-3\cos^2\theta_q) \nonumber \\ 
&+& A^f_4(Y, M, Q_T) \cos\theta_q \Big\} \, ,
\end{eqnarray}
}

\noindent 
where $Y$, $M$ and $Q_T$ are the rapidity, invariant mass and transverse momentum of the $Z$ boson. The index $f$ 
denote the flavor, and the term $\alpha_f$ is proportional to the cross section of the quark-antiquark subprocesses (with flavor $f$). 
The terms $A^f_0$ and $A^f_4$ 
are the normalized polar angular coefficient functions, while other angular functions are cancelled when integrated over the azimuthal angle. 
Both $A^f_0$ and $A^f_4$, as a function of $Y$, $M$ and $Q_T$, can be calculated in the perturbative expansion of the electroweak and QCD 
couplings. In this work, they are computed using the {\sc ResBos}~\cite{resbos} package at approximate next-to-next-to-leading order (NNLO) 
plus next-to-next-to-leading logarithm (NNLL) in QCD interaction.

The canonical scales~\cite{QCDpaper1, QCDpaper2} are set to the invariant mass of the lepton pair in the DY events. The EW calculation of 
{\sc ResBos} is similar to the effective born approximation used in {\sc zfitter}~\cite{ZFITTER}.
The DY events are categorized as forward ($F$) for $\cos\theta_q>0$, and backward ($B$) for $\cos\theta_q<0$. It is custom to 
define $A_{FB}$ as 
\begin{eqnarray}
A_{FB} = \frac{\sigma_F - \sigma_B}{\sigma_F + \sigma_B}
\end{eqnarray}
where $\sigma_F$ and $\sigma_B$ are the cross sections of $F$ and $B$ events. In this case, the parton densities of the initial state quarks 
and antiquarks contribute only to $\alpha_f$ terms, and $A^f_0$, $A^f_4$ and $A_{FB}$ are independent of proton structure information. 
However, such $A_{FB}$ cannot be experimentally observed because the directions of $q$ and $\bar{q}$ are practically unknown.

At the LHC, the DY process is observed in terms of $\cos\theta_h$, of which $H_A$ is defined as the hadron which points to the same hemisphere 
aligned to the reconstructed $Z$ boson of the dilepton final state. Since the $Z$ boson is boosted along with the direction of the parton 
which carries higher energy, $\cos\theta_h = \cos\theta_q$ when $q$ has larger energy than $\bar{q}$, and $\cos\theta_h = -\cos\theta_q$ 
when $\bar{q}$ has the larger energy. 
We introduce the dilution factor $D_f(Y, M, Q_T)$ to represent the probability of having $\cos\theta_h = -\cos\theta_q$, {\it i.e.,} 
the antiquark has higher energy than the quark in the hard scattering subprocess. In this manner, the differential cross section in term 
of $\cos\theta_h$ can be expressed as:

\begin{footnotesize}
\begin{eqnarray}
\label{eq02:dsigmadh}
 & &\frac{d\sigma}{d\cos\theta_h dY dM dQ_T}= \frac{3}{8} \sum_f \alpha_f(Y, M, Q_T) \nonumber \\
  &\times&  \left\{ (1+\cos^2\theta_h) + \frac{1}{2}A^f_0(Y, M, Q_T)(1-3\cos^2\theta_h) \right. \nonumber \\ 
    &+& \left. [1-2D_f(Y, M, Q_T)]A^f_4(Y, M, Q_T) \cos\theta_h \right\}
\end{eqnarray}
\end{footnotesize}

\noindent It is clear that only the $A^f_4$ term would be affected by the dilution effect. Accordingly, the experimental observed asymmetry 
$A^h_{FB}$ can be formulated as follow:

\begin{footnotesize}
\begin{eqnarray}
\label{eq03:fullfactorize}
& &A^h_{FB}(Y, M, Q_T) = \nonumber \\
& & \frac{\sum_f [1-2D_f(Y, M, Q_T)]\alpha_f(Y, M, Q_T) A^f_{FB}(Y, M, Q_T)}{\sum_f \alpha_f(Y, M, Q_T)}
\end{eqnarray}
\end{footnotesize}

\noindent 
where $A^f_{FB}(Y, M, Q_T)$ denotes the asymmetry calculated using $\cos\theta_q$ for each $f\bar{f}$ subprocess, which is independent with 
proton structure.  

The derivation of Eq.~\ref{eq03:fullfactorize} is straightforward and provided in Sec.~S1 of the Supplemental Material~\cite{SupMat}.
The above equation reveals the nature of the observed $A^h_{FB}$,which is an average of the asymmetry contributed from each $f\bar{f}$ 
subprocess, weighted by its relative cross section, 
and the associated dilution factor reflects our limited knowledge of the $q$ and $\bar{q}$ PDFs of the proton.
 
Here, we assume that the parton distributions of strange quark and antiquark of the proton are the same, $s(x)=\bar{s}(x)$, though a small 
violation can be generated at the NNLO via DGLAP parton evolution~\cite{DGLAP}. Similarly, $c(x)=\bar{c}(x)$ and $b(x)=\bar{b}(x)$, with $c$ 
and $b$ denoting charm and bottom flavor, respectively. Consequently, the dilution factors $D_s$, $D_c$ and $D_b$ equal to 0.5 in all $Z$ 
boson kinematic configurations, and their contributions in the numerator of Eq.~\ref{eq03:fullfactorize} vanish.

Based on the argument, we propose to factorize the observed $A^h_{FB}$ as products of the proton structure parameters and independent EW-dominant 
parts:

\begin{footnotesize}
\begin{eqnarray}
\label{eq04:finalfactorize1}
&& A^h_{FB}(Y, M, Q_T) = \nonumber \\
& & [\Delta_u(Y, M, Q_T) + P^u_0(Y, Q_T)] \cdot A^u_{FB}(Y, M, Q_T; \effstw) + \nonumber \\
& & [\Delta_d(Y, M, Q_T) + P^d_0(Y, Q_T)] \cdot A^d_{FB}(Y, M, Q_T; \effstw) \,,
\end{eqnarray}
\end{footnotesize}

\noindent where the structure parameters are related to  Eq.~\eqref{eq03:fullfactorize} via 

\begin{footnotesize}
\begin{eqnarray}\label{eq05:finalfactorize2}
P^u_0(Y, Q_T) &=& \int  \frac{[1-2D_u(Y, M, Q_T)] \alpha_u(Y, M, Q_T)}{\sum_f \alpha_f(Y, M, Q_T)} dM \Big/ \int dM \nonumber \\ 
P^d_0(Y, Q_T) &=& \int  \frac{[1-2D_d(Y, M, Q_T)] \alpha_d(Y, M, Q_T)}{\sum_f \alpha_f(Y, M, Q_T)} dM \Big/ \int dM \nonumber \\ 
\Delta_u(Y, M, Q_T) &=& \frac{[1-2D_u(Y, M, Q_T)] \alpha_u(Y, M, Q_T)}{\sum_f \alpha_f(Y, M, Q_T)}  - P^u_0(Y, Q_T) \nonumber \\ 
\Delta_d(Y, M, Q_T) &=& \frac{[1-2D_d(Y, M, Q_T)] \alpha_d(Y, M, Q_T)}{\sum_f \alpha_f(Y, M, Q_T)} - P^d_0(Y, Q_T) 
\nonumber \\
\end{eqnarray}
\end{footnotesize}

\noindent 
In Eq.~\ref{eq04:finalfactorize1}, the $\effstw$-dependence of the light quark asymmetry $A^{u/d}_{FB}(Y, M, Q_T; \effstw)$ are explicitly 
written out, which can be precisely predicted by the SM electroweak calculation. The structure parameters $P^{u/d}_0(Y, Q_T)$ represent 
the parton information averaged in a given mass range of $M$, while the parameters $\Delta_{u/d}(Y, M, Q_T)$ describe the variation around 
the averaged behavior provided by $P^u_0$ and $P^d_0$, respectively, in that mass range. 
Both $P^f_0$ and $\Delta_f(M)$ parameters are structure parameters, representing the parton information. Since $A^h_{FB}$ is usually observed 
in a narrow mass window around the $Z$ pole, the relevant proton structure information and the corresponding uncertainties are dominated by 
$P^f_0$ parameters. 

\begin{figure}[!hbt]
\begin{center}
\epsfig{scale=0.4, file=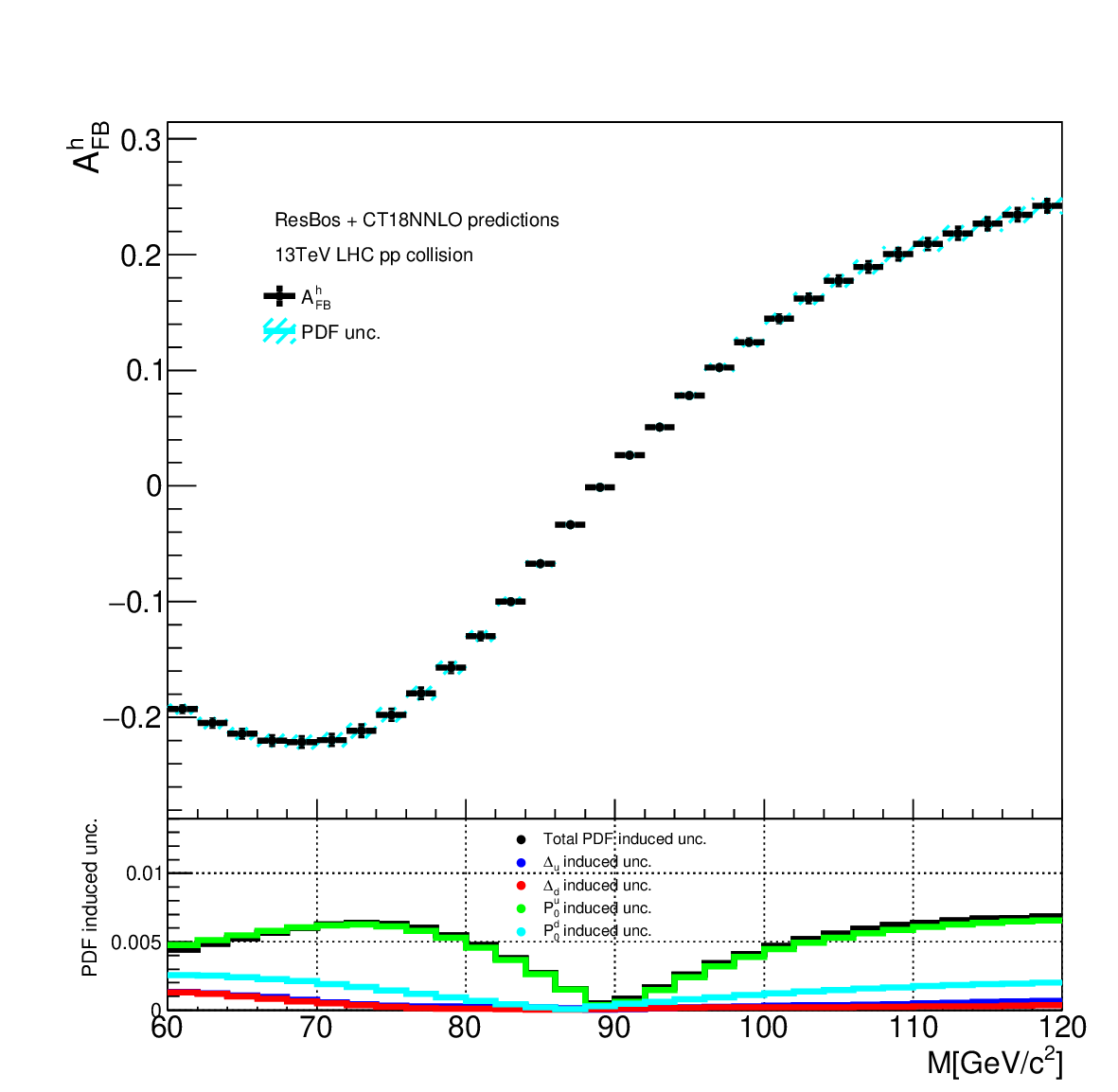}
\caption{\small The spectrum of $A^h_{FB}(M)$, and the corresponding uncertainties from $P^u_0$, $P^d_0$, $\Delta_u(M)$ and $\Delta_d(M)$ terms, 
respectively. The central values and of $A^h_{FB}$ and the uncertainties in each $M$ bin are predicted by CT18 PDFs. The uncertainties correspond 
to $68\%$ C.L. 
The observables are averaged over $Y$ and $Q_T$ in this figure.
}
\label{fig01:AFBvsM}
\end{center}
\end{figure}

As an example, we show in the upper panel of Figure~\ref{fig01:AFBvsM} the distribution of $A^h_{FB}(M)$, in the mass region 
of $60 < M < 120$ GeV/$c^2$, for a pseudo-data sample of the DY process produced at the 13 TeV LHC, generated using the {\sc ResBos} program 
with CT18 NNLO PDFs~\cite{CT18PDF}. In the lower panel of Figure~\ref{fig01:AFBvsM}, the PDF-induced uncertainty of $A^h_{FB}$ for each 
invariant mass bin is decomposed according to Eq.~\ref{eq04:finalfactorize1}, for illustration. As expected, the uncertainty is dominated 
by those of $P^u_0$ and $P^d_0$, while the contributions from $\Delta_u(M)$ and $\Delta_d(M)$ are quite small. 

\begin{figure}[!hbt]
\begin{center}
\epsfig{scale=0.4, file=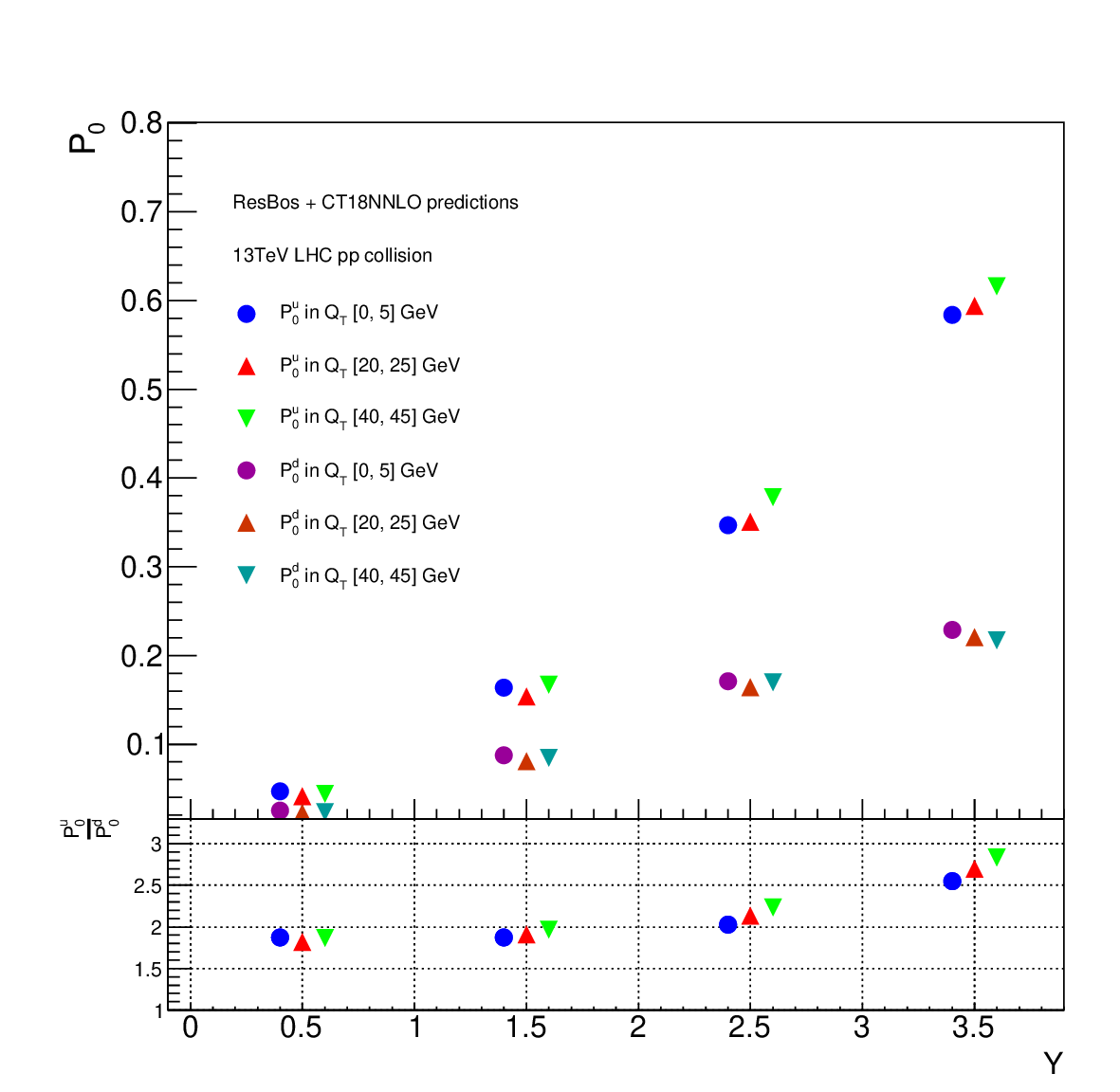}
\caption{\small The $P^u_0$ and $P^d_0$ values as a function of $Q_T$ and $Y$ predicted by {\sc ResBos} + CT18. For the $Y$ dependence, 
results in the plots correspond to the $P_0$ values averaged over $|Y|$ in $[0, 1.0]$, $[1.0, 2.0]$, $[2.0, 3.0]$ and $[3.0, 4.0]$ bins. 
For the $Q_T$ dependence, the $P_0$ values are averaged in three regions of [0, 5], [20, 25] and [40, 45] GeV, as examples.
}
\label{fig02:P0vsZY}
\end{center}
\end{figure}

By its definition in Eq.~\ref{eq05:finalfactorize2}, the $P^f_0$ parameters contain information of the dilution effects of the light quarks and 
their relative cross sections. Accordingly, the values of $P^f_0$ are expected to be increasing as a function of $Y$, for the dilution effects 
should be reduced in the kinematic region where the difference in the quark and anti-quark energy is large. 
At the LHC energies, both $(1-2 D_u)$ and $(1-2 D_d)$ approach to 0 when $Y$ is zero, and toward 1 as the magnitude of $Y$ increases. This trend 
is demonstrated in Figure~\ref{fig02:P0vsZY}, where the $P^f_0$ values predicted by CT18 PDFs are shown as a function of $Y$ and $Q_T$. 
Additionally, in the high $Y$ region where the dilution effects of the $u$ and $d$ quarks are similarly small, the separation of $P^u_0$ and $P^d_0$ 
could be attributed to the difference in the light quark parton abundance and their corresponding cross sections. 
Note that if the ratio of $P^u_0$ and $P^d_0$ is employed, the information on $s$, $c$ and $b$ cross sections cancels out. Therefore, the experimental 
observation on such ratio can directly probe the ratio of $u$ and $d$ (anti)quark parton densities of the proton as a function of $x$. 
These phenomenal conclusions are also checked using predictions of MSHT20~\cite{MSHT20} and NNPDF3.1~\cite{NNPDF31}, respectively, as shown in 
Sec.~S2 of the Supplemental Material.

\section{Simultaneous fit of $\effstw$ and $P^f_0$}

Some progresses on reducing the PDF and EW correlation and potential bias have been made in the past. In Ref.~\cite{CPCstudy}, it was found that 
excluding the $A^h_{FB}$ data around the $Z$ pole region in a PDF global analysis could reduce the bias. To further reduce the bias, at the expense 
of sensitivity, a new observable was proposed later in Ref.~\cite{AFBshape}, which uses the gradient information of $A^h_{FB}(M)$ spectrum 
to perform the PDF global fitting, while using its average value around the $Z$-pole to determine $\effstw$. However, the residual bias given by 
the above two methods is still sizable.
Alternatively, Ref.~\cite{NuisanceFit} explored the impact of the imperfect knowledge of the proton structure on the determination of EW parameters 
(e.g. $M_W$) via nuisance parameter formalism, by including the bin-by-bin correlation of the kinematic distributions with respect to PDF variations. 
Though the analysis would effectively reduce the PDF uncertainty, it relies on the information of specific PDF sets, which are not automatically 
updated in the process of measuring the EW parameter.
In short summary, none of these analyses provided a strategy to simultaneously fit $\effstw$ and parton information.

Based on the factorization of Eq.~\ref{eq04:finalfactorize1}, we propose a new method to simultaneously determine 
the EW and proton structure parameters by fitting to the observed $A^h_{FB}(Y, M, Q_T)$ data. 
One can employ Eq.~\ref{eq04:finalfactorize1} to build theoretical 
templates, with $\effstw$, $P^u_0(Y, Q_T)$ and $P^d_0(Y, Q_T)$ as fitting parameters. Their values are therefore determined by requiring the best 
agreement between theory templates and the observed data. 
Due to lack of sufficient constraints by using the $A^h_{FB}$ distribution only, 
the $\Delta_f(Y, M, Q_T)$ terms would have to be fixed to the prediction of current PDFs, and thus induce 
extra theoretical uncertainties to the fitted parameters. 

\begin{figure}[!hbt]
\begin{center}
\epsfig{scale=0.4, file=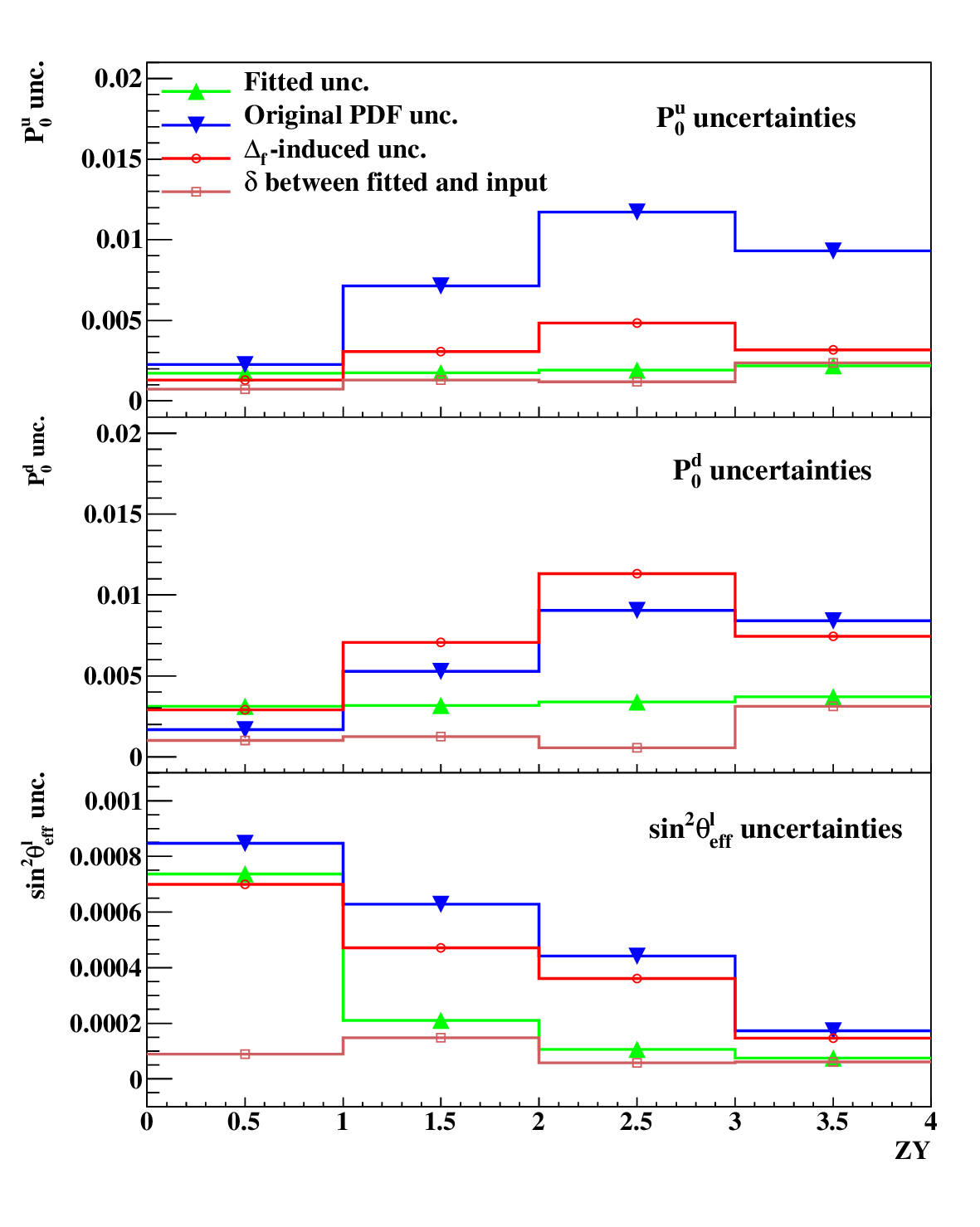}
\caption{\small 
Various uncertainties of $P^u_0$ (upper), $P^d_0$ (middle) and $\effstw$ (lower), as detailed in the main text. 
Results are given in the $Q_T$ range of [5, 10] GeV, as an example.
}
\label{fig03:fittedP0}
\end{center}
\end{figure}

For numerical test, a {\sc ResBos}+CT18 pseudo-data sample of 2.5 billion DY events is used, corresponding to 
1 $ab^{-1}$ integrated 
luminosity at the LHC 13 TeV pp collisions. 
Similar tests are also made using MSHT20 and 
NNPDF3.1, of which the results are given in Sec. S3 of the Supplemental Material.
The fitted results of $P^u_0$, $P^d_0$ and $\effstw$ parameters, as a 
function of $Y$, are depicted in Figure~\ref{fig03:fittedP0}. 
The statistical fluctuations of the fitted $P^u_0$, $P^d_0$ and $\effstw$ 
values (labelled as ``Fitted unc.''), are comparable to the difference between 
the fitted values and their input or predicted values in the pseudo-data 
(labelled as ``$\delta$''), which manifests the closure of the factorization and the 
fitting method. 

The theoretical uncertainties associated with the input $\Delta_f$ terms in the fit (labelled as ``$\Delta_f$-induced 
unc.'') are estimated by repeating the fitting procedure with different $\Delta_f(Y, M, Q_T)$ predictions given by 
the CT18 PDF error sets, instead of the central set, with $P^f_0$ and $\effstw$ as free fitting parameters, and 
quoting the variations of their fitted values, respectively. For comparison, the variations of $P^f_0$ values predicted 
by different CT18 PDF error sets, and the PDF-induced uncertainty on the $\effstw$ extracted from the 
convnentional method of setting $\effstw$ as the only fitting parameter, are also depicted (labelled as ``Original PDF unc.''), 
respectively. The factorization method provides a new perspective for the issue of the correlation between PDFs and $\effstw$. 
Firstly, the $\Delta_f$-induced uncertainties arise from the variation of proton structure information in a small region 
of the Bjorken variable $x$, while the original PDF ones are dominated by the average magnitude of the 
structure parameters. Secondly, we note that the $\Delta_f$ terms do not contribute large uncertainty on 
$A^h_{FB}$ distribution itself, cf. Figure~\ref{fig01:AFBvsM}. However, when $A^h_{FB}$ is the sole data included 
in the fit, the $P^f_0$ terms and $\effstw$ are higly (negatively) correlated. To improve the resolution power 
of the proposed factorization form of $A^h_{FB}$ to the determination of the $P^f_0$ terms and $\effstw$, 
we could either introduce new $\Delta_f$-sensitive observables in the PDF global fitting to reduce the 
uncertainty on $\Delta_f$ itself, or add additional $P^f_0$-sensitive data in the simultaneous fit, 
so that the correlation between $\effstw$ and $P^f_0$ can be reduced. Of course, this kind of new 
$\Delta_f$ or $P^f_0$-sensitive observables, preferably at the LHC, ought to be $\effstw$-independent.

\section{Conclusion}

The $A^h_{FB}$ measurement at hadron colliders not only is an ideal observable for the determination of $\effstw$, but also provides unique proton 
structure information on the relative difference between quarks and antiquarks. It is essential to factorize the proton structure information 
with well defined observables, so that both $\effstw$ and the parton information can be simultaneously determined. 
In this article, we proposed a novel method to do just that, which is based on the factorization property of Eq.~\eqref{eq04:finalfactorize1},
and demonstrated how the method works. 
The $A^h_{FB}$ factorization provides a novel method fo handle the proton distribution 
uncertainties in the electroweak measurement at the LHC, in contrast to the conventional PDF error estimation. 
It should also be pointed out that by introducing other observables in the same LHC data, which are sensitive to 
either of the two types of factorized $\Delta_f$ or $P^f_0$ terms, the precision of the 
simultaneous fit based on the $A^h_{FB}$ factorization could be further improved, so that the proton 
structure information and the EW $\effstw$ parameter could be determined model-independently in the 
future LHC experiment.

~\\

\section{Acknowledgements}
This work was supported by the National Natural Science Foundation of China under Grant No. 11721505, 11875245, 
12061141005 and 12105275, and supported by the ``USTC Research Funds of the Double First-Class Initiative''. 
This work was also supported by the U. S. National Science Foundation under Grant No. PHY-2013791. 
C.-P. Yuan is also grateful for the support from the Wu-Ki Tung endowed chair in particle physics.

\end{document}